\documentclass[twocolumn,prl,superscriptaddress,preprintnumbers,amsmath,amssymb]{revtex4}

\usepackage{graphicx}
\usepackage{dcolumn}
\usepackage{bm}
\usepackage{float}
\usepackage[]{color}

\begin{document}

\title{Direct imaging of orbitals in quantum materials}

\author{Hasan~Yava\c{s}} 
\altaffiliation{Present address: SLAC National Accelerator Lab., 2575 Sand Hill Rd, Menlo Park, CA 94025, USA}
\affiliation{Max Planck Institute for Chemical Physics of Solids, N{\"o}thnitzer Stra{\ss}e 40, 01187 Dresden, Germany}
\affiliation{PETRA III, Deutsches Elektronen-Synchrotron (DESY), Notkestra{\ss}e 85, 22607 Hamburg, Germany}
\author{Martin~Sundermann}
 \affiliation{Max Planck Institute for Chemical Physics of Solids, N{\"o}thnitzer Stra{\ss}e 40, 01187 Dresden, Germany}
 \affiliation{Institute of Physics II, University of Cologne, Z{\"u}lpicher Stra{\ss}e 77, 50937 Cologne, Germany}
\author{Kai~Chen}
	\altaffiliation[present address: ]{Synchrotron SOLEIL, L'Orme des Merisiers, Saint-Aubin, BP~48, 91192 Gif-sur-Yvette Cedex, France}
  \affiliation{Institute of Physics II, University of Cologne, Z{\"u}lpicher Stra{\ss}e 77, 50937 Cologne, Germany}
\author{Andrea~Amorese}
	\affiliation{Max Planck Institute for Chemical Physics of Solids, N{\"o}thnitzer Stra{\ss}e 40, 01187 Dresden, Germany}
  \affiliation{Institute of Physics II, University of Cologne, Z{\"u}lpicher Stra{\ss}e 77, 50937 Cologne, Germany}
\author{Andrea~Severing}
	\affiliation{Max Planck Institute for Chemical Physics of Solids, N{\"o}thnitzer Stra{\ss}e 40, 01187 Dresden, Germany}
  \affiliation{Institute of Physics II, University of Cologne, Z{\"u}lpicher Stra{\ss}e 77, 50937 Cologne, Germany}
\author{Hlynur Gretarsson}
	\affiliation{Max Planck Institute for Chemical Physics of Solids, N{\"o}thnitzer Stra{\ss}e 40, 01187 Dresden, Germany}
	\affiliation{PETRA III, Deutsches Elektronen-Synchrotron (DESY), Notkestra{\ss}e 85, 22607 Hamburg, Germany}
\author{Maurits~W.~Haverkort}
 \affiliation{Institute for Theoretical Physics, Heidelberg University, Philosophenweg 19, 69120 Heidelberg, Germany}
\author{Liu~Hao~Tjeng}
 \affiliation{Max Planck Institute for Chemical Physics of Solids, N{\"o}thnitzer Stra{\ss}e 40, 01187 Dresden, Germany}

\begin{abstract}
The spectacular physical properties of quantum materials based on transition metal, rare earth, and 
actinide elements continue to challenge our comprehension of solid state physics and chemistry. The electronic 
states of these materials are dominated by the $d$ and $f$ wave functions intertwined with the strong 
band formation of the solid. In order to estimate which wave functions contribute to the ground state formation, we have had to rely, until now, on theoretical calculations combined with spectroscopy. Here we show that $s$-core-level non-resonant 
inelastic x-ray scattering ($s$-NIXS) can directly image the active orbital in real space, without the necessity of any modeling. The power and accuracy of this new technique is shown using the text-book example, x$^2$-y$^2$/3$z^2$-r$^2$ orbital of the Ni$^{2+}$ ion in NiO single crystal. 
\end{abstract}

\maketitle

The search for new materials with novel properties is commonly focused on materials containing transition metal, rare-earth, and/or actinide elements. The presence of the atomic-like $d$ or $f$ wave functions provide a fruitful playground to generate novel phenomena\,\cite{Cava,Khomskii2014,Keimer2015,Wirth2016,Pfleiderer2009}. The intricate interplay of band formation with the local electron correlation and atomic multiplet effects leads to phases that are nearly isoenergetic, making materials' properties highly tunable by doping, temperature, pressure, or magnetic field. Understanding the behavior of the $d$ and $f$ electrons is essential for designing and controlling novel quantum  materials. Therefore, identifying the $d$ or $f$ wave functions that actively participate in the formation of the ground state is crucial. So far, these wave functions have been mostly deduced from optical,  x-ray and neutron spectroscopy methods in which spectra must be analyzed and interpreted using theory or modeling. This, however, is also a challenge in and of itself since \textit{ab-initio} calculations hit their limits due to the many-body nature of the problem. Here, we have established an experimental method that circumvents the need for involved analysis, and instead, provides the information as measured. With this technique, we can make a direct image of the active orbital and determine what the atomic-like object looks like in a real solid.

The spectral intensity of the dipole-allowed $s$\,$\rightarrow$\,$p$ 
transition depends on the orientation of the electric field polarization vector of the photon relative to the orientation 
of the $p$ orbital\,\cite{Fowles1968}. Since the $s$ orbital is spherically symmetric, sweeping the polarization vector over all angles yields an angular intensity distribution that directly maps the shape and orientation of the $p$ orbital hole. Yet, material research requires knowledge of $d$ and/or $f$ orbital shapes.
As the $s$\,$\rightarrow$\,$d$ or $s$\,$\rightarrow$\,$f$ transitions are dipole forbidden, it has been challenging to develop an experimental method that has non 
vanishing matrix elements \textsl{beyond the dipole limit}. However, the relatively new experimental method of non-resonant inelastic x-ray scattering (NIXS), available due to modern synchrotron facilities with high brilliance, has offered new potential.

The interaction of light with matter is given by two terms: a term proportional to the scalar product of the electron momentum operator $\vec{p}$ and the photon vector potential $\vec{A}$, and a term proportional to the vector potential $\vec{A}$ squared. When photon energy matches an atomic resonance, the $\vec{p}$\,$\cdot$\,$\vec{A}$ term dominates; off-resonance, the interaction is governed by the $\vec{A}$$^2$ term. Focusing on this last term using NIXS, the double
differential cross-section $\frac{d^2\sigma}{d\Omega d\omega}$ becomes proportional to the dynamical structure factor S($\vec{q}$,$\omega$)\,\cite{Schulke2007} which contains the material-specific information we are seeking:
\begin{eqnarray}
S(\vec{q},\omega) = \sum_{f}|\left\langle f|e^{i\vec{q} \cdot \vec{r}}|i \right\rangle |^2 \delta(\hbar\omega_i-\hbar\omega_f-\hbar\omega),\nonumber
\end{eqnarray}
where $|i \rangle$ and $|f \rangle$ denote the (many-body) initial and final states, $\vec{q}$\,=\,$\vec{k_i}$\,-\,$\vec{k_f}$ the transferred momentum, $\hbar\omega$\,=\,$\hbar\omega_i$\,-\,$\hbar\omega_f$ the transferred energy, and $\vec{k}$$_{i,f}$ and $\hbar\omega_{i,f}$ the momentum and energy of the incoming and scattered photons, respectively.

Beyond-dipole matrix elements appear in the scattering cross-section when expanding the transition operator 
e$^{i\vec{q} \cdot \vec{r}}$ to the $k^{th}$ order, whereby $k$ denotes the \textsl{multipole order of the 
scattering cross-section}\,\cite{Schulke2007,Haverkort2007,Gordon2008,Gordon2009,Bradley2010,
Caciuffo2010,Bradley2011,Laan2012,Laan2012b,Willers2012}. The so-called triangular condition and parity rule 
restrict the number of multipoles to $|l_f-l_i|$\,$\le$\,$k$\,$\le$\,$l_f+l_i$ and $|l_i$\,+\,$l_f$\,+\,$k|$\,=\,even 
for a $l_i$\,$\rightarrow$\,$l_f$ transition (respective orbital momenta of initial and final state). This implies 
that for a $d$\,$\rightarrow$\,$f$ transition, only dipole ($k$\,=\,1), octopole ($k$\,=\,3), and triakontadipole 
($k$\,=\,5) scattering orders occur, and for $s$\,$\rightarrow$\,$d$ -a dipole-forbidden transition- only 
the quadrupole transition with $k$\,=\,2 contributes to S($\vec{q},\omega$). 
For small momentum transfers $|\vec{q}|$, the NIXS spectra very much resemble dipole-allowed x-ray absorption spectroscopy (XAS). In other words, the NIXS $2p/3p$\,$\rightarrow$\,$3d$ excitations in transition metal compounds\,\cite{Haverkort2007,Gordon2009} 
or the $3d/4d$\,$\rightarrow$\,$4f/5f$ and $5d$\,$\rightarrow$\,$5f$ excitations in rare earth and actinide materials\,\cite{Gordon2008,Gordon2009} exhibit line shapes that are very similar to the ones obtained from XAS\,\cite{Chen1992,deGroot1994,Tanaka1994,Csiszar2005,Hansmann2008}. The only difference is that in NIXS, 
the direction of the momentum transfer $\hat{q}$ ($\vec{q}$/$|\vec{q}|$) provides the information that is 
obtained from the electrical vector polarization in XAS. However, for large $|\vec{q}|$, for the same metal ion, 
the $p$\,$\rightarrow$\,$d$ or $d$\,$\rightarrow$\,$f$ transitions yield a different spectral distribution with 
additional features that cannot be seen in a dipole-based XAS experiment\,\cite{Schulke2007,Haverkort2007,Gordon2008,Gordon2009,Bradley2010,Caciuffo2010,Bradley2011,Laan2012,Laan2012b,Willers2012,Sundermann2016,Sundermann2018}. Moreover, the dipole-forbidden $s$\,$\rightarrow$\,$d$ or $s$\,$\rightarrow$\,$f$ transitions (quadrupolar or octopolar, respectively) now have non-vanishing matrix elements, and consequently are allowed and become visible.

\begin{figure}[]
	\centering
	 \includegraphics[width=0.98\columnwidth]{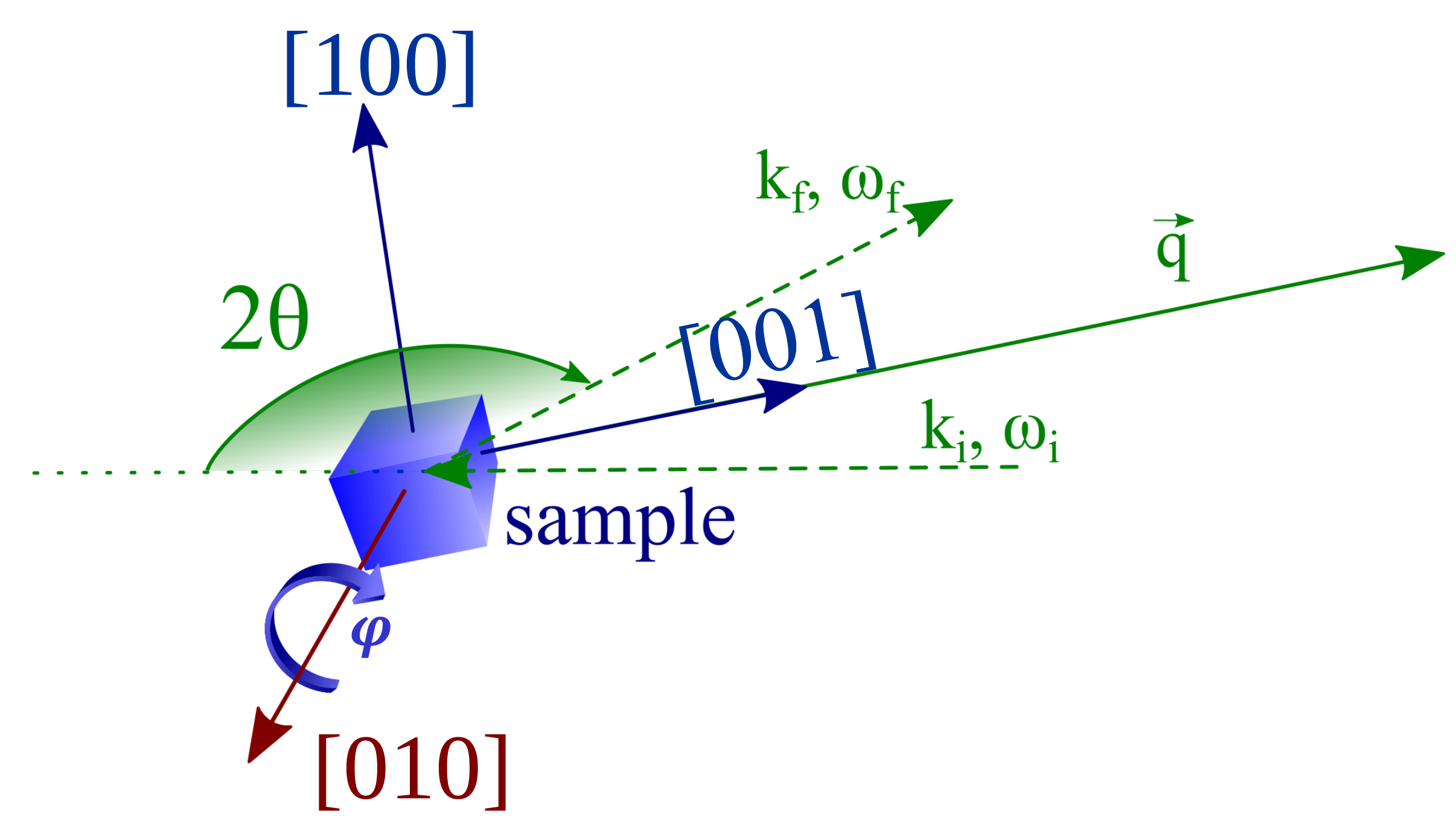}
		\caption{The scattering geometry is defined by the incoming and the scattered beam, $k_i,\omega_i$ and $k_f,\omega_f$, 
respectively (dashed green arrows). This geometry, which outlines the scattering triangle, remains fixed throughout 
the measurements. The single crystal sample (blue) is rotated around an axis perpendicular to the scattering plane 
(maroon) by an angle $\varphi$, and for each $\varphi$ an inelastic spectrum is collected. Here, $\varphi$\,=\,0 refers 
to $\vec{q}$$\|$[001] (specular geometry).}
		\label{fig1}
\end{figure}

The novelty of our approach is to exploit these $s$-core-level transitions involving our search for a new method to determine - quantitatively and model free - the local valence orbitals that make up the electronic structure of $d$ and $f$ containing quantum materials. We investigated the $s$\,$\rightarrow$\,$d$ transition in an inelastic x-ray scattering experiment ($s$-NIXS) at large momentum transfers $|\vec{q}|$ and map the quadrupolar scattering intensity as a function of the direction of the momentum transfer $\hat{q}$ relative to the crystal lattice. We used a single crystal of NiO, an antiferromagnetic insulator\,\cite{Sawatzky1984}, with a Ni $d^8$ configuration as a model system; and the large momentum transfers were guaranteed by high scattering angles and hard x-rays.

In our experimental setup as illustrated in \verb+Fig.+\,1, S($\vec{q}$,$\omega$) of the NiO sample was 
recorded as a function of the sample angle $\varphi$, here defined as the angle between the fixed momentum 
transfer vector $\vec{q}$ and the NiO surface normal (see Methods). \verb+Fig.+\,2 shows a compilation 
of NIXS spectra measured for many different sample angles. The spectra show the M$_{2,3}$ edge 
($3p$\,$\rightarrow$\,$3d$) of nickel at around 70\,eV and, most importantly, the dipole-forbidden
M$_1$ ($3s$\,$\rightarrow$\,$3d$) excitations at around 110\,eV, overlaid on the broad Compton profile.
The signal to background ratio is excellent in the energy range of the M$_1$ edge. A close-up of this
edge and its directional dependence on $\vec{q}$ along $\vec{q}$$\|$[001] and $\vec{q}$$\|$[100]
are displayed in \verb+Fig.+\,3\,a, and for $\vec{q}$ along $\vec{q}$$\|$[001] and $\vec{q}$$\|$[110]
in \verb+Fig.+\,3\,b\,and\,c. In the close-up plots, the Compton profile has been subtracted using a simple linear
background (see Methods).

\begin{figure}[]
	\centering
	 \includegraphics[width=0.98\columnwidth]{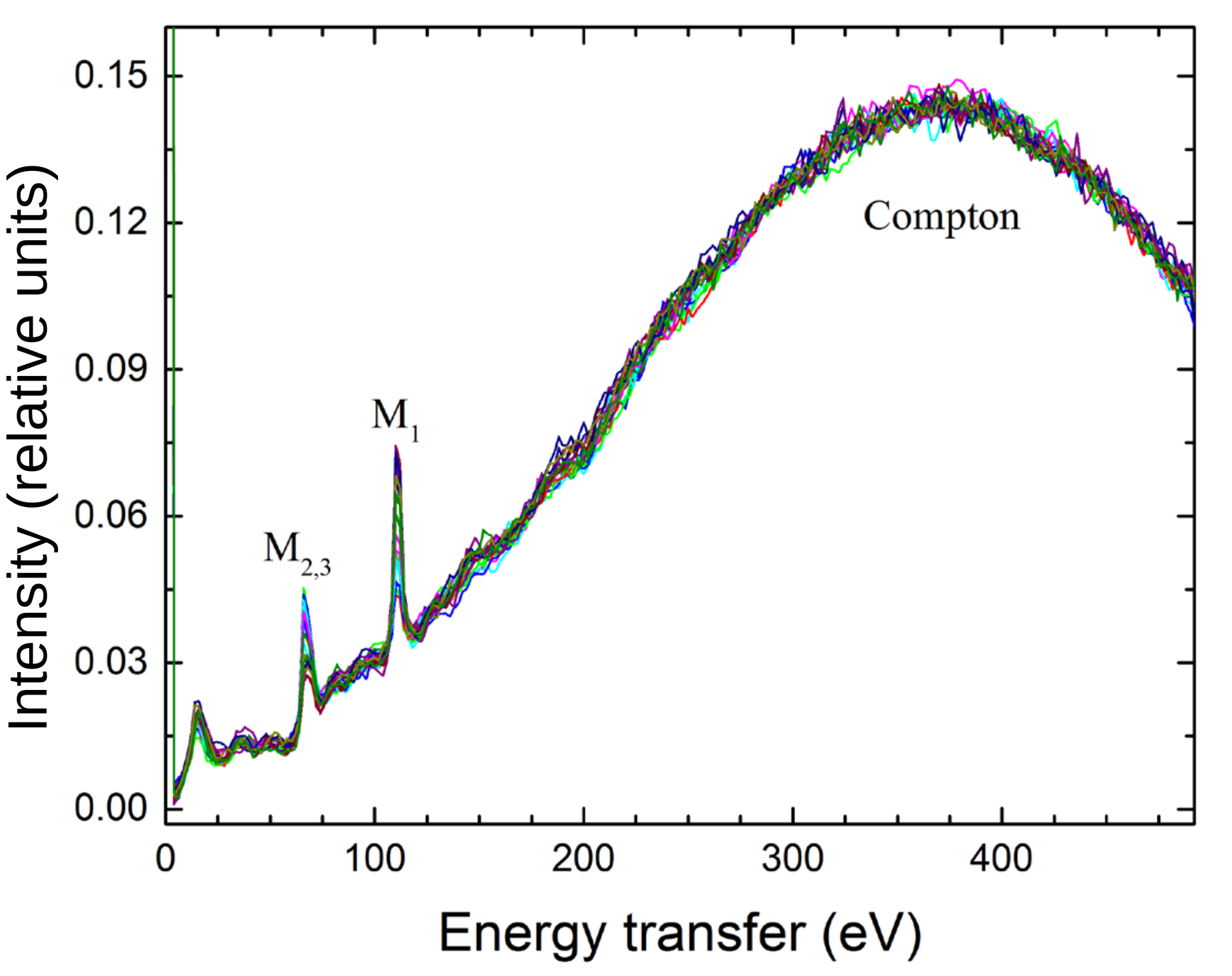}
		\caption{Experimental NIXS spectra of NiO. 
Compilation of spectra, which were collected for a variety of crystal rotations $\varphi$ with respect to the 
geometrically fixed momentum transfer vector $\vec{q}$. The graph shows all the spectra. The Compton 
profile peaks at approximately 350\,eV and is used for data normalization. The dipole-allowed Ni 
M$_{2,3}$ ($3p$\,$\rightarrow$\,$3d$) edge at around 70\,eV energy transfer and the dipole-forbidden 
Ni M$_1$ ($3s$\,$\rightarrow$\,$3d$) excitations at around 110\,eV can be clearly observed. The data exhibit 
an excellent signal to background (Compton) ratio in the energy range of the Ni edges, rendering NIXS as 
a high-contrast experiment.}
		\label{fig2}
\end{figure}

\begin{figure*}[]
	\centering
	 \includegraphics[width=1.5\columnwidth]{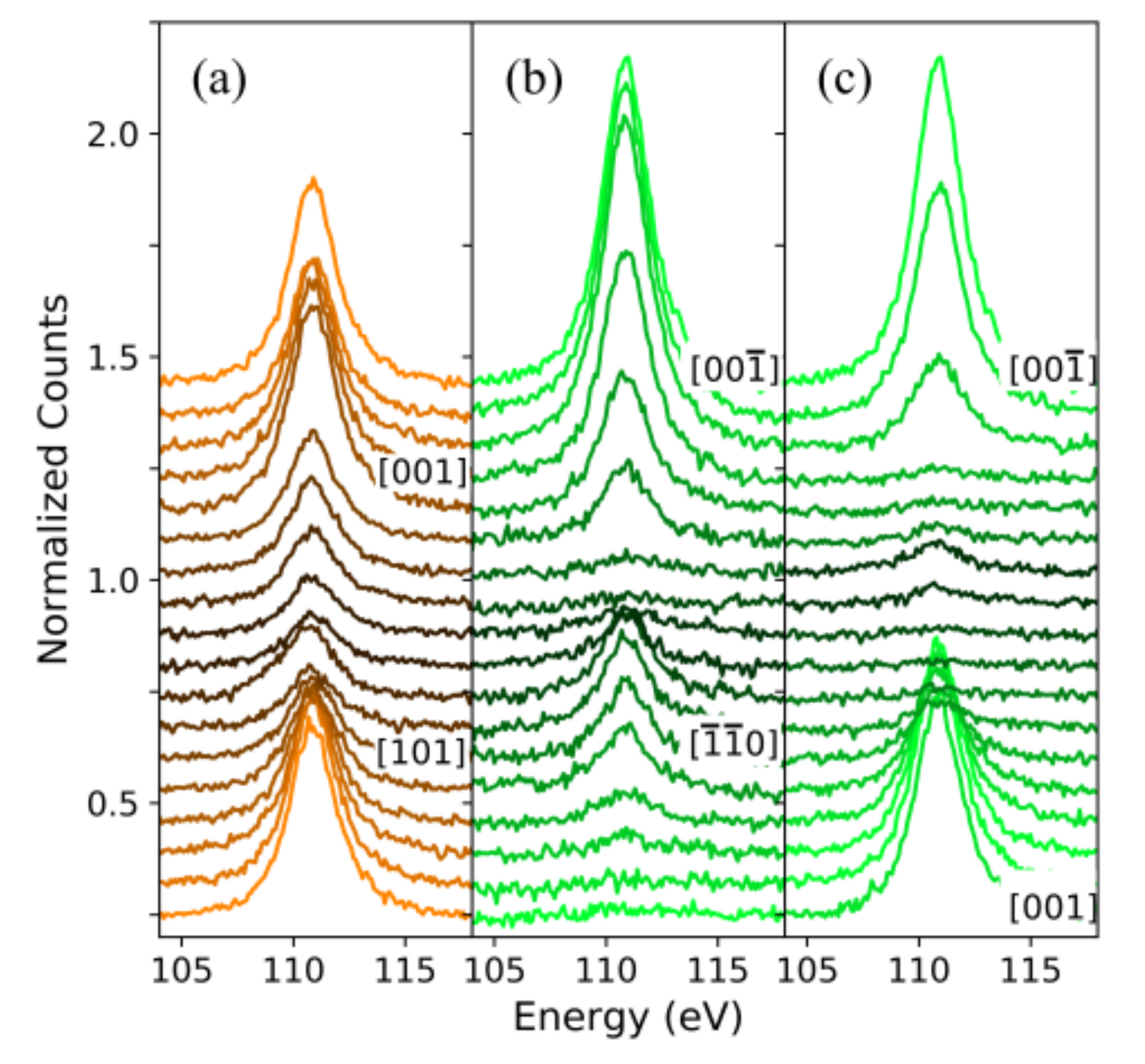}
		\caption{Ni $M_1$ ($3s$\,$\rightarrow$\,$3d$) edge spectra. 
A close-up view of the relevant range of the spectra after background subtraction and normalization. 
Panel (a) shows data for momentum transfer vector $\vec{q}$ sweeping on a plane defined by $\vec{q}$$\|[001]$ and $\vec{q}$$\|[100]$
and (b-c) on a plane defined by $\vec{q}$$\|[001]$ and $\vec{q}$$\|$[110]. Spectra corresponding to major axes are marked accordingly, and the spectra in between are plotted sequentially. Panel (b) shows the spectra, where $\vec{q}$ is sweeping clockwise from $\vec{q}$$\|[00\bar{1}]$, passing $\vec{q}$$\|[\bar{1}\bar{1}0]$, and ending where the projection of the orbital function vanishes. 
Similarly, panel (c) shows data, where $\vec{q}$ is sweeping counterclockwise from $\vec{q}$$\|[00\bar{1}]$ towards $\vec{q}$$\|[001]$. The data are vertically 
shifted for clarity. The scattering intensity of the M$_1$ edge excitations ($3s$\,$\rightarrow$\,$3d$) 
depends strongly  on the relative orientation of momentum transfer and 
crystallographic direction in the NiO single crystal.}
		\label{fig3}
\end{figure*}

\begin{figure*}[]
	\centering
	 \includegraphics[width=1.98\columnwidth]{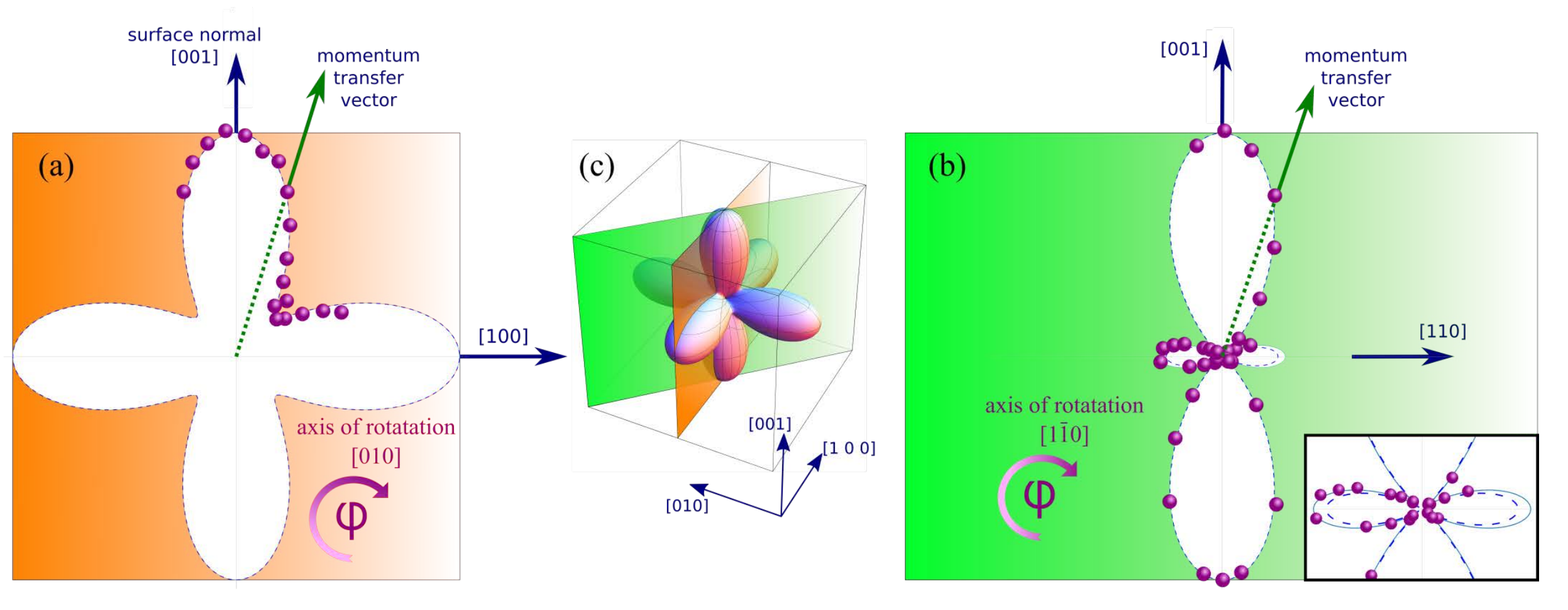}
		\caption{Orbital shape of $^3A_2$ $3d$($x^2$\,-\,$y^2$)$3d$($3z^2$\,-\,$r^2$) hole density. 
Three-dimensional hole density distribution of the Ni high-spin 3$d^8$ configuration \textbf{(c)}. Projection of 
the three-dimensional orbital shape on two planes defined by \textbf{(a)} $[001]$ and $[100]$ and \textbf{(b)} 
$[001]$ and $[110]$. Data points on polar plots \textbf{(a-b)} are integrated intensities for Ni $M_1$ 
($3s$\,$\rightarrow$\,$3d$) for corresponding $\varphi$, which is the angle between the momentum transfer 
vector $\vec{q}$ and the surface normal vector $[001]$ for both cases \textbf{(a-b)}. For \textbf{(a)} the 
sample is rotated such that the $\vec{q}$ sweeps between $[001]$ and $[100]$, and for \textbf{(b)} 
$\vec{q}$ sweeps between $[001]$ and $[110]$. Inset in \textbf{(b)} demonstrates the theoretical orbital function (blue dashed line) and the corrected function as a result of an angular convolution with the 3x4 analyzer array. For \textbf{(a)}, the correction was insignificant.}
		\label{fig4}
\end{figure*}

To quantitatively analyze the $3s$\,$\rightarrow$\,$3d$ transition's directional dependence,
we determined the integrated intensity of each spectrum in \verb+Fig.+\,3  and placed it on a polar plot 
as displayed in (\verb+Fig.+\,4). Panel\,(a) shows the data points for $\vec{q}$ sweeping in the 
[001]--[100] plane (orange), and panel\,(b) for $\vec{q}$ in the [001]--[110] plane (green). They fall 
accurately on top of the orbital shapes which denote 'cuts' through the [001]--[100] (orange) and [001]--[110] 
(green) planes of the three-dimensional orbital hole state of the Ni high-spin $3d^8$ configuration in
octahedral coordination, namely the $^3A_2$ $3d$($x^2$\,-\,$y^2$)$3d$($3z^2$\,-\,$r^2$) as
shown in \verb+Fig.+\,4\,(c). This means that, for the first time, we have generated a purely experimental method that can directly visualize the fundamental atomic-like quantum mechanical objects in solids. The information that we have obtained is extremely detailed; for example, we can see nicely and clearly see the small lobes of the $3d$($3z^2$\,-\,$r^2)$. Here we note that we have used two orbital shapes in \verb+Fig.+\,4\,(b): 
the blue dashed  line is the $3d$($x^2$\,-\,$y^2$)$3d$($3z^2$\,-\,$r^2$) function and the solid
line is the same function convoluted with the angular acceptance of the 3x4 analyzers we used in our
experiment (see Methods in Section Supplementary). The near perfect agreement further demonstrates the accuracy of the method.

The directional dependence of the integrated $s$-NIXS intensity at the Ni $M_1$ edge
($3s$\,$\rightarrow$\,$3d$) directly maps the local orbital hole of the ion in the ground state.
There is no need to carry out multiplet analysis of the spectral line shape to extract this information,
in contrast to, for example, the non-$s$ edges (e.g. $L_{2,3}$($2p$), $M_{2,3}$ ($3p$), $M_{4,5}$ ($3d$),
$N_{4,5}$ ($4d$), and $O_{4,5}$ ($5d$)) in both NIXS\,\cite{Gordon2008,Gordon2009,Bradley2010,
Caciuffo2010,Bradley2011,Willers2012,Schulke2007,Haverkort2007,Sundermann2016,Sundermann2018} 
and XAS experiments\,\cite{Chen1992,deGroot1994,Tanaka1994,Csiszar2005,Hansmann2008}.
The reason is fundamental: The $M_1$ ($3s$\,$\rightarrow$\,$3d$)
quadrupolar excitation process involves a spherically symmetric $s$ orbital, so the angular distribution of
intensity is solely determined by the hole charge distribution in the initial state with respect to the
momentum transfer $\vec{q}$. This is similar to the dipole-allowed $s$\,$\rightarrow$\,$p$ transition in XAS,
where an angular sweep of the polarization dependence maps out the orientation of the $p$ hole directly.
We would like to emphasize that details of the $s$-NIXS final states do not matter because the information
is extracted from the integrated intensity of the spectra (i.e. from the sum of the intensities of all final states).
As a result, only the properties of the initial state are probed. This is analogous to using spectral sum 
rules to extract expectation values of the relevant quantum numbers of the system in the ground state\,\cite{Csiszar2005,Thole1992,Carra1993}. The power of $s$-NIXS, as compared to XAS, is that it allows 
transitions not only from $s$-to-$p$, but also from $s$-to-$d$ and $s$-to-$f$ due to the possibility of 
going beyond the dipole limit when using large momentum transfers $\left|\vec{q}\right|$.

The $s$-NIXS process involves a core hole, meaning that both the electronic structure of the system and consequently the measured valence hole are projected locally.
The intensity distribution is not what would be measured in an 
x-ray diffraction (XRD) experiment, even if such an experiment could be carried out with sufficient accuracy. In fact, it would  be extremely difficult for transition metal, rare earth, and actinide compounds to be measured with the desired 
accuracy in XRD due to their relatively small number of valence electrons with respect to core electrons. 
$s$-NIXS provides information complementary to that from an XRD experiment by elucidating which local orbital 
or atomic wave function is active.

The $s$-NIXS method presented here is not limited to ionic materials. In cases where configuration interaction
effects play an important role due to covalency or itineracy, the image of the probed local orbital 
will reflect these effects directly. The strength of $s$-core-level NIXS is that the information
is extracted from the $\vec{q}$-directional dependence of the \textit{integrated} intensity and not from the
\textit{line shape} of the spectra. Thus, the details of the final states are no longer important, rendering complex
configuration interaction calculations unnecessary. The sole $\vec{q}$-directional dependence is rooted in the spherical symmetry of the $s$-core hole.

To conclude, we have directly imaged one of the fundamental quantum mechanical objects in crystals,
namely, a $d$-orbital, which is derived purely mathematically from first principles. We have revealed that 
non-resonant inelastic x-ray scattering involving an $s$-core level is an extremely powerful and accurate
experimental method to determine the local orbital in the ground state. Albeit low cross-section, the excellent 
signal to background ratio allows for highly reliable results. The procedure relies on the integrated intensity of 
the signal, which tremendously simplifies the interpretation as there is no need to carry out multiplet analysis 
of the spectral line shape to extract the desired information. The method is element specific, which is invaluable 
for unraveling the different origins for electron correlation effects in complex materials. Since the probing 
photons have high penetrating power, the measurements are bulk sensitive, and can be performed with 
complex sample environments (e.g. small samples, high pressures, high/low temperatures). We believe that 
this method opens up new opportunities for the study of a wide range of $d$ and $f$ electron containing 
materials, where knowledge of their local wave function is of central importance to reveal their underlying 
physics and thus provide guidance for the design of new quantum materials.

\section{Acknowledgment} M.\,S., K.\,C., A.\,A. and A.\,S. gratefully acknowledge support from the German funding agency DFG under Grant No SE1441-4-1. 

\section{Appendix}
\subsection{Experiment:}Non-resonant inelastic x-ray scattering (NIXS) measurements were performed at the High-Resolution Dynamics Beamline P01 of PETRA-III synchrotron in Hamburg, Germany. \verb+Fig.+\,1 illustrates the experimental setup, showing the incoming beam ($\vec{k}$$_i,\omega_i$), sample, scattered beam ($\vec{k}$$_f,\omega_f$), and the corresponding momentum transfer vector ($\vec{q}$). The energy of the x-ray photon beam incident on the sample was tuned with a Si(311) double-reflection crystal monochromator (DCM). The photons scattered from the sample were collected and energy-analyzed by an array of twelve  spherically bent Si(660) crystal analyzers. The analyzers are arranged in a 3x4 configuration. The energy of the analyzers ($\hbar\omega_f$) was fixed at 9690\,eV; the energy loss spectra were measured by scanning the energy of the DCM ($\hbar\omega_i$). Each analyzer signal was individually recorded by a position-sensitive custom-made LAMBDA detector. The energy calibration was regularly checked by measuring the zero-energy-loss position of each spectrum. The best possible energy resolution was guaranteed by pixel-wise analysis of the detector recordings and measured as 0.7\,eV (FWHM).

The positioning of the analyzer array determines the momentum transfer vector and the corresponding scattering triangle, which is defined by the incident and scattered photon momentum vectors, $\vec{k}$$_i$ and $\vec{k}$$_f$, respectively. The large scattering angle (2$\theta$\,$\approx$\,155$^{\circ}$) chosen for the current study assured a large momentum transfer of $\left|\vec{q}\right|$ = (9.6\,$\pm$\,0.1)\,\AA$^{-1}$ when averaged over all analyzers. $\vec{k}$$_f$ and 2$\theta$ were kept constant by fixing the energy and the position of the analyzer array. Since the energy transfer range of interest (100 to 120\,eV) was small with respect to the incident and final energies ($\sim$9700\,eV), variation of $\vec{k}$$_i$ during energy scanning was insignificant. This guaranteed that the scattering triangle was virtually unchanged throughout the course of the experiment with $\left|\vec{q}\right|$\,$\approx$\,constant.

\subsection{Sample} NiO single crystal (SurfaceNet, Germany) was kept at T\,=\,20K throughout the experiment. 
It was aligned as in \verb+Fig.+\,1 and rotated by angle $\varphi$ around an axis perpendicular 
the [010] lattice direction; $\vec{q}$$\|$[001] ($\varphi$\,=\,0) corresponds to specular geometry. 
Energy scans were taken for many values of $\varphi$ so that the directional dependence of 
S($\vec{q}$,$\omega$) could be measured for $\vec{q}$ sweeping between $\vec{q}$$\|$[001] and $\vec{q}$$\|$[100]. For the second set of measurements, the crystal was reoriented to evaluate 
S($\vec{q}$,$\omega$) on the $\vec{q}$$\|$[001]-$\vec{q}$$\|[110]$. This time, the axis 
of rotation was along $[1\bar{1}0]$.

\subsection{Data Treatment} The data were normalized to the Compton peak at about 350 \,eV energy transfer (see \verb+Fig.+\,2). Subsequently, a linear background was subtracted from each spectrum in order to account for the Compton scattering in the energy range of the Ni M$_1$ edge.

Since the analyzer array (3x4) is spread over a finite solid angle, each measured spectrum S($\vec{q}$,$\omega$) includes an array of momentum transfer vectors $\vec{q}$ corresponding to individual analyzers. In this case, taking an average $\vec{q}$ does not work for directions where the orbital wave function varies significantly for small angular changes (i.e. small lobes of the $3d$($3z^2$\,-\,$r^2)$). The theoretical orbital wave function should be convoluted with the angular spread of the analyzer array to reflect this effect. The inlet on \verb+Fig.+\,4\,(b) demonstrates the theoretical function (blue dashed line) and the convoluted function (solid line), which agrees well with the data points.


\end{document}